\documentclass[a4paper,12pt]{article}

\usepackage{float}
\usepackage{amsmath}
\usepackage{amssymb}
\usepackage{graphicx}
\usepackage{graphicx}
\usepackage{lscape}
\usepackage{epic}

\begin{document}


\title{Controlling self-organized criticality in complex networks}
\author{Daniel O. Cajueiro$^{1, 3}$ and Roberto F. S. Andrade$^{2, 3}$}
\date{}
\maketitle

\begin{center}$^1$Department of Economics -- Universidade de Bras\'{i}lia, DF 70910-900, Brazil.\\$^2$Instituto de F\'{i}sica, Universidade Federal da Bahia, BA 40210-340,
Brazil.\\$^3$National Institute of Science and Technology for
Complex Systems,  Brazil.\end{center}

\begin{abstract} A control scheme to reduce the size of avalanches of
the Bak-Tang-Wiesenfeld model on complex networks is proposed.
Three network types are considered: those proposed by
Erd\H{o}s-Renyi, Goh-Kahng-Kim, and a real network representing
the main connections of the electrical power grid of the western
United States. The control scheme is based on the idea of
triggering avalanches in the highest degree nodes that are near to
become critical. We show that this strategy works in the sense
that the dissipation of mass occurs most locally avoiding larger
avalanches. We also compare this strategy with a random strategy
where the nodes are chosen randomly. Although the random control
has some ability to reduce the probability of large avalanches,
its performance is much worse than the one based on the choice of
the highest degree nodes. Finally, we argue that the ability of
the proposed control scheme is related to its ability to reduce
the concentration of mass on the network.
\end{abstract}

\section{Introduction}

The phenomenon of self-organized criticality (SOC) on complex
networks~\cite{bonabeau,lispac02,goh03,goh04,goh05} has recently
been studied in order to understand the failures that take place
in real networks such as electric power distribution and internet
~\cite{goh03}. Attempts for analytical approaches of SOC behavior
on geometrically grown networks have also been
reported~\cite{vie07}. A review of these attempts may be found for
instance in reference~\cite{dor08}. The main idea is that, due to
the strong relation among the neighbors, a small overload that
occurs in a node or a small collection of nodes may spread into
the whole network. Parallel to this, there is a growing literature
that deals with the issue of robustness and the reduction of
overload failure cascades caused by removal of nodes (attacks or
overload) in complex
networks~\cite{hol02,mot04,alb04,cru04,zhao05,hua08}. The remedy
often used in order to avoid the propagation of these cascades is
the intentional removal of some special nodes characterized by
their degree or by a given centrality measure.

In this paper, we propose a control scheme to reduce the
probability of large avalanches in a generalization of the
Bak-Tang-Wiesenfeld (BTW) sandpile model~\cite{bak87} to complex
network substrates ~\cite{goh03,goh04,goh05}. In order to provide
a motivation to this problem, consider the situation in which
there is a demand for limited recourses in each node of a complex
network. Since the resources are limited, if the demand in one
node exceeds a given threshold, a demand avalanche happens: the
local node resource provider in this node is closed and the demand
is forwarded to its neighbors. Being a SOC system, it is clear
that a local demand avalanche in one node may be amplified to the
node's neighbors, transforming small events into large ones.
However, large avalanches are undesired in the system, since they
may destabilize several resource providers simultaneously. One way
of avoiding this kind of phenomenon is to close, for a short time
lapse, the resource provider in the node and to move the demand to
its neighbors. The main idea is that instantaneous closures can
avoid big avalanches. The difficult is how to choose the correct
moment and place of the such closure.

The issue of controlling SOC in regular lattices, where mass is
added and removed from a system, has been recently discussed in
sandpile models~\cite{cajand10}. In that paper, we have shown that
an external control action, which amounts to triggering avalanches
in sites that are near to be come critical, was able to reduce the
probability of very large events, so that mass dissipation occurs
most locally. Due to the homogeneity of the lattice where
traditional SOC phenomena have been investigated, one difficult
present in~\cite{cajand10} is that, in order to make the decision
whether an avalanche should or not be exogenously triggered, one
had to simulate a replica model of the region of the system to be
controlled. Here, differently from~\cite{cajand10}, the control
scheme does not depend on the replica model and, therefore, is
less costly than the one presented in~\cite{cajand10}.
Furthermore, while in~\cite{cajand10} we were interested in
controlling the size of avalanches in only a region of the system,
in this paper we are interested in controlling the size of the
avalanches in the whole network.

The strategy considered here to control the size of avalanches in
BTW systems may not be confused with the stock exchange trading
halts~\cite{sub94,lee94}. While our strategy allows that the
demands be attended in another provider, in such circuit breakers
the traders are not able to meet their demands for a finite time
lag. Although our work is related to network robustness
literature, in the sense that we pay a special attention to some
nodes based on their centrality, it is very different in essence.
Here, we do not remove the nodes of the network nor the demand
associated to the node -- the control rule keeps the mass of the
system on being the same. The only variation of mass in the BTW
system is due to the SOC dynamics, which injects it at a low
constant rate and remove it in some specific sites. Furthermore,
different from these works, our work trigger small avalanches in
order to avoid large ones.

Our work is also related to the literature of immunization of
complex networks~\cite{sat02,coh03,gal07,che08} where the
strategies for immunization are based on the specific properties
of the nodes of the network. However, very different from them, we
consider the issue of controlling self organized system without
the immunization (changing the state of the node) of nodes.

The control scheme we propose is based on triggering avalanches on
a percentage of the highest degree nodes. We compare this strategy
with another cheaper control scheme, based on triggering
avalanches on randomly selected nodes with the same frequency used
in the first scheme. While the former strategy clearly requires
global knowledge of the network, the latter does not need this
information. Although both strategies are able to reduce the size
of avalanches, the former performs much better. We also show that
the problem of reducing the size of avalanches of SOC systems on
complex networks is related to the reduction of the concentration
of mass on few nodes of the system. Finally, we report the cost of
the control scheme based on the percentage of nodes that the
control scheme has to deal with and number of interventions that
the control scheme performs in a given time window $N_T$.

\section{Controlled BTW model in complex networks} The BTW sandpile
model~\cite{bak87} has been recently studied
~\cite{goh03,goh04,goh05} in an scale-free network discussed by
Goh, Kahng and Kim (GKK)~\cite{goh01}. It has also been studied in
the random Erd\H{o}s-Renyi (ER) networks ~\cite{bonabeau}. Here,
we follow closely a previously developed approach
~\cite{goh03,goh04,goh05}, but study the problem for GKK and ER
networks, as well as for the network that represents the actual
electrical power grid of the western United
States~\cite{watstr98}.

Consider a network with $n$ nodes. Let $k(i)$ be the degree of
node $i\in\{1,\cdots,n\}$ and $\mathcal{N}(i)$ be the set of
neighbors of node $i$. Assume also that each node
$i\in\{1,\cdots,n\}$ stores a certain amount $z_{i}$ of mass
units. The dynamics of the BTW model in complex networks may be
described by the following two rules: (a) Addition rule: at each
time step, a mass unit is added to a randomly selected node
$i\in\{1,\cdots,n\}$, so that $z_{i}\rightarrow z_{i}+1$. (b)
Toppling rule: if $z_{i}\ge z_{ic}=k(i)$, then $z_{i}\rightarrow
z_{i}-k(i)$, $z_{j}\rightarrow z_{j}+1, \forall j\in
\mathcal{N}(i)$.

In order to control the BTW model, we propose here the so-called
Highest Degree Nodes Control Based (H-control), which assumes that
we have global knowledge of the network. The idea is to choose the
percentage of highest degree nodes $p_{H}$ of the network that
will be controlled and to build a set $\mathcal{S}_{H}$ with these
nodes. To be controlled here means that if $z(i)=z_c(i)-1$ of a
node $i\in \mathcal{S}_{H}$ in a given time, the control system
causes this node to topple. We will use the term {\it explosion}
to identify such action in order to differentiate it from the
ordinary addition rule. As one may see further, there are in fact
two technical differences from this process to the usual addition
rule: (1) The mass of the system is not increased; (2) Since the
site is not critical, the neighbors do not receive the same amount
of mass.  The avalanche that follows from the explosion caused by
the intervention of the control scheme will be called as {\it
controlled avalanches}, so that it will be possible to distinguish
them from the {\it uncontrolled avalanches} that happen due to the
deposition of mass in SOC dynamics. Since the controlled avalanche
is triggered by emptying the non critical node $i$, the available
mass in this site goes randomly to some of the neighbors belonging
to $\mathcal{N}(i)$. This means there is a pre-defined order
inside the set of highest degree nodes. This is interesting since
starting from the highest degree nodes in this set, it is possible
that some explosions that could be necessary before the beginning
of the control intervention may not be necessary anymore. This can
happen if a node that belongs to the set of highest degree nodes
is also connected to a note that has higher degree than it. We
compare this control scheme with the so-called Random Selected
Nodes Control Based (R-control), which assumes no information
about the network, but it intervenes with the same frequency of
the H-control. In each instant of time, it selects randomly the
nodes to be controlled and triggers an explosion on this node, if
$z(i)=z_c(i)-1$. Both control schemes assume that we keep the same
mass in the system. An easier strategy, which is available in real
world problems, is simply reduce the mass of the system. In this
case, instead of triggering explosions in the nodes, one could
transport the excess of mass from the system. Since we think that
this strategy modify strongly the dynamics of the system, we do
not consider it in this paper. However, an interesting issue that
can be studied, is how to optimize the choice of the mass that
should be transported away -- since there is also a high cost
associated to the transportation of mass.

\section{Results}
As advanced in the previous section, we have applied this control
scheme to BTW model on three network types. The GKK network can be
built based on the following algorithm \cite{goh01}: start with
$n$ nodes $i\in\{1,\cdots,n\}$ and assign to each of them a weight
equal to $w_i=i^{-\alpha}$, where $\alpha\in [0,1]$ is related to
the degree exponent according to $\gamma=1+1/\alpha$. Then select
two different nodes $i,j\in\{1,\cdots,n\}$ with probability equal
to the normalized weights $w_i/\sum_{k=1}^{n} w_k$ and
$w_j/\sum_{k=1}^{n} w_k$, respectively, and connect them if they
are not already connected. The exponent of the avalanche size
distribution in these GKK networks was determined to be
$\tau=\gamma/(\gamma-1)$~\cite{goh03}. The ER networks were built
assuming that all nodes of the network have the same constant
probability $p$ of being connected. The exponent of the avalanche
size distribution in the networks was determined to be
$\tau=1.5$~\cite{bonabeau}.

\begin{figure}
  \includegraphics[width=8cm,height=8cm]{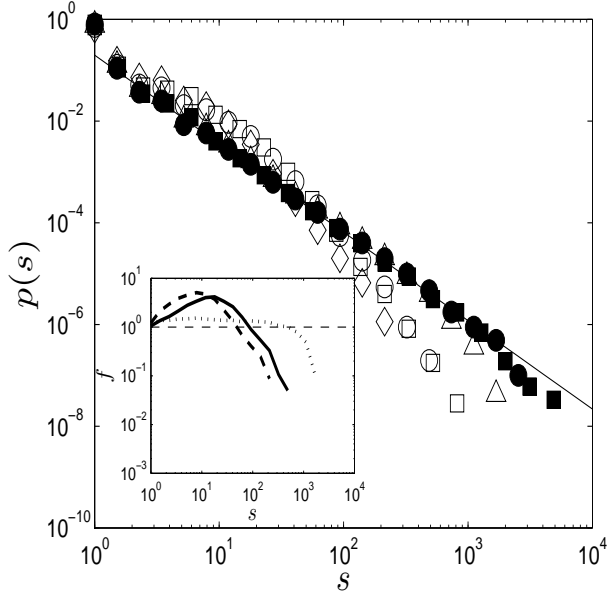}
\caption{Probability distribution of avalanche sizes $p(s)$ in the
GKK network with mean average degree 4. Points were obtained by
logarithmic size bins over the whole range of $s$. Solid and
hollow symbols denote uncontrolled and controlled system,
respectively. Uncontrolled system sizes $n=10^5$ (squares) and
$10^4$ (circles). Symbol types indicate the following values of
$(n, \mathrm{control\; scheme\; (H\; or\; R)},  p_{H}, N_T)$ for
the controlled systems: squares ($10^5$, H, $5\%$, 0.27), circles
($10^4$, H, $5\%$, 0.25), diamonds ($10^4$, H, $10\%$, 0.52) and
triangles ($10^4$, R, -- , 0.25). In the inset, curves for the
ratio $f$ between total number of avalanches in the controlled and
uncontrolled simulations. The number of time steps are equal for
both simulations. Line types are as follows. Scale-free network:
solid $(10^4, \mathrm{H}, 5\%, 0.25)$, dashes $(10^4, \mathrm{H},
10\%, 0.52)$ and dots $(10^4, \mathrm{R}, 5\%, 0.25)$. The curve
for ($10^5$, H, $5\%$, 0.27) was not shown since it is difficulty
to differentiate this curve from the solid one. } \label{figura1}
\end{figure}

\begin{figure}
  \includegraphics[width=8cm,height=8cm]{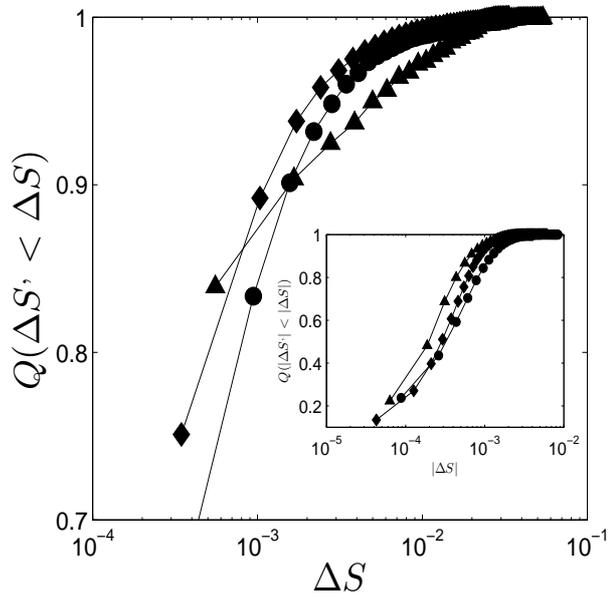}
\caption{Complementary probability $Q(\Delta S^{,} < \Delta S)$
for positive values of $\Delta S^{,}$ when $n=10^4$. Symbol types
indicate the following values of $(\mathrm{control\; scheme\;(H\;
or\; R)}, p_{H}, N_T)$: circles $(\mathrm{H}, 5\%, 0.25)$,
diamonds $(\mathrm{H}, 10\%, 0.52)$ and triangles $(\mathrm{R}, -
, 0.25)$. In the inset, complementary probability of $Q(|\Delta
S^{,}| < |\Delta S|)$ for negative values of $\Delta S^{,}$ with
the same legend.}\label{figura2}
\end{figure}

\begin{figure}
  \includegraphics[width=8cm,height=8cm]{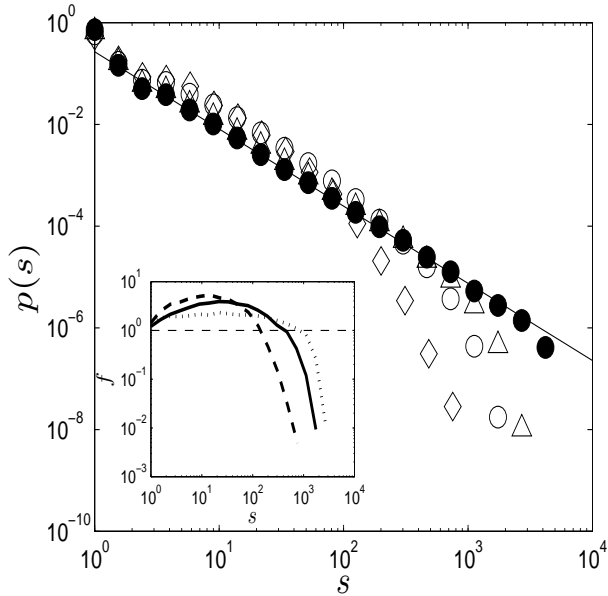}
\caption{Simulations for a ER network with the same mean average
degree of the GKK network presented in Fig.1. Probability
distribution of avalanche sizes $p(s)$ in the the ER network.
Points were obtained by logarithmic size bins over the whole range
of $s$. Solid denotes the uncontrolled system. Hollow Symbol types
indicate the controlled system with the following values of
$(\mathrm{control\;scheme\; (H\; or\; R)}, p_{H}, N_T)$: circles
$(\mathrm{H}, 2.5\%, 0.42)$, diamonds $(\mathrm{H},5\%$, 0.76) and
triangles $(\mathrm{R}, - , 0.50)$. In the inset, curves for the
ratio $f$ between total number of avalanches in the controlled and
uncontrolled simulations. The number of time steps are equal for
both simulations. Line types indicate the following values of
$(\mathrm{control\; scheme\;(H\; or\; R)}, p_{H}, N_T)$: solid
$(\mathrm{H}, 2.5\%, 0.42)$, dashes $(\mathrm{H}, 5\%, 0.76)$ and
dots $(\mathrm{R}, - , 0.50)$.} \label{figura3}
\end{figure}

\begin{figure}
  \includegraphics[width=8cm,height=8cm]{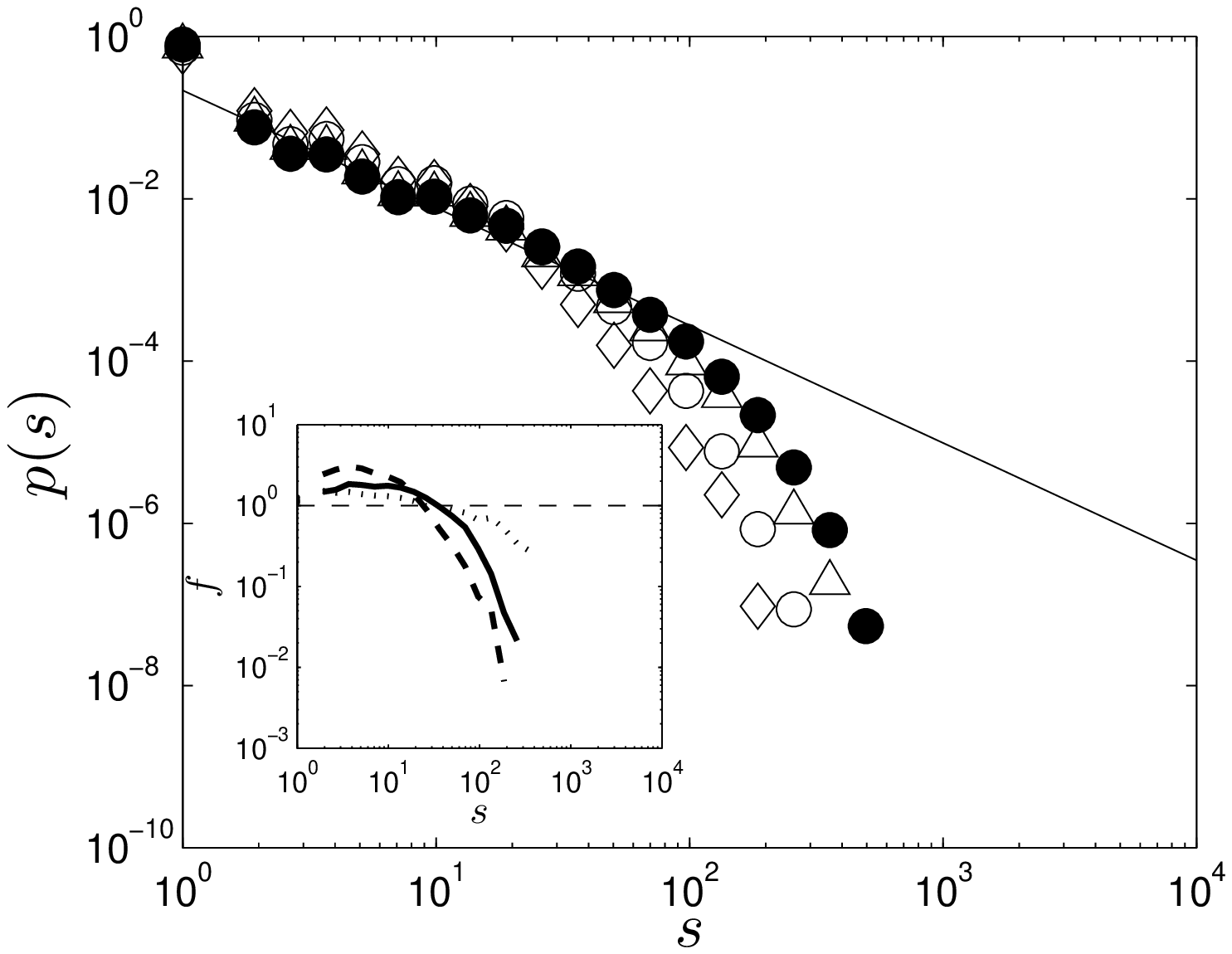}
\caption{Probability distribution of avalanche sizes $p(s)$ in the
the electrical power grid of the western United States with
$n=4941$. Points were obtained by logarithmic size bins over the
whole range of $s$. Solid denotes the uncontrolled system. Hollow
Symbol types indicate the controlled system with the following
values of $(\mathrm{control\;scheme\;(H\; or\; R)}, p_{H}, N_T)$:
circles $(\mathrm{H}, 5\%, 0.24)$, diamonds $(\mathrm{H},10\%$,
0.55) and triangles $(\mathrm{R}, - , 0.25)$. In the inset, curves
for the Ratio $f$ between total number of avalanches in the
controlled and uncontrolled simulations. The number of time steps
are equal for both simulations. Line types are as follows: solid
$(\mathrm{H}, 5\%, 0.24)$, dashes $(\mathrm{H}, 10\%, 0.55)$ and
dots $(\mathrm{R}, - , 0.25)$.} \label{figura4}
\end{figure}

Fig.\ref{figura1} compares the the probability distribution
function (PDF) of avalanche sizes $p(s)$ of the uncontrolled
system (solid symbols) with that of the system controlled by the
H-control (hollow symbols). While the data of the uncontrolled
system include only the uncontrolled avalanches, those of the
controlled system include controlled and uncontrolled avalanches.
The PDF of the degrees of the nodes of the GKK networks presented
in this figure is a power-law with theoretical exponent
$\gamma=3.0$, since we used the value of $\alpha=0.5$ to build
them. For both networks with $10^4$ and $10^5$ nodes we have
numerically obtained the exponent $\hat \gamma=3.07$. The straight
line in Fig. \ref{figura1} is the best fit to the data for the
systems with size $n=10^4$ and $10^5$ in the interval
$s\in[10^{0.3},10^{3.3}]$. Based on this data, the exponent of the
avalanche size distribution in these GKK networks was determined
to be $\hat \tau=1.74$ which is roughly the value of the
empirically one determined in~\cite{goh03}. Fig.\ref{figura1} also
shows that the H-control is able to strongly reduce the
probability of large events. Besides, this figure shows that when
$p_{H}$ increases, the control scheme is more efficient in the
reduction of large size avalanches. Finally, we also compare the
efficiency of H-control scheme with R-control scheme. We can see
that although both are efficient to reduce the probability of
large avalanches, the former is much more efficient. The decrease
in the probability of large avalanches results from the fact that,
since only saturated sites are exploded by random process, some of
them sites are correctly chosen. Furthermore, since the most
connected nodes have by definition more neighbors, even if the
explosion are wrongly selected, these explosions are likely to
have some effect in the most connected nodes.

In the inset of Fig.\ref{figura1} we evaluate the efficiency of
the control system by the ratio $f$ between the number of
avalanches of the controlled to the uncontrolled system. It makes
clear that the control system is actually reducing the number of
large size avalanches, i.e., its effect is not restricted to
increasing the number of small and medium size events. In this
inset, we also illustrate the effect of increasing $p_{H}$. It is
intuitive that, if $p_{H}$ is decreased, the controller is less
efficient to reduce the chance of large avalanches, but smaller
$p_{H}$ are clearly more economical. This can be seen in the small
$s$ region of the inset of Fig.\ref{figura1}, where the number of
small avalanches of the controlled system with $p_{H}=0.05$ is
smaller than that with $p_{H}=0.1$. $p_{H}$ plays a role similar
to the acceptable size $a_c$ considered in~\cite{cajand10}. Two
costs are of relevance in this control scheme, namely the cost of
scanning the high degree nodes in order to see if they are
saturated and the cost of the intervention (explosion). Both of
them increases linearly with $p_{H}$.

We have also performed simulations considering GKK networks with
other values of $\alpha$ (such as $\alpha=0.66$) implying in
avalanches with different exponents of power law size
distributions. Although there is a variation in the setting of the
control scheme and in the number of interventions, the control
schemes perform much like the ones presented in Fig.\ref{figura1}.

An intuitive issue that was also checked in simulations is that,
in the uncontrolled SOC system, the highest degree nodes always
saturate in huge avalanches. Besides accumulating a large amount
of mass, the highest degree nodes are also able to distribute a
large amount of mass along the network. It is worth mentioning
that the other nodes of the network clearly also accumulate mass
and distribute that mass along the network in a scale proportional
to their degrees. Therefore, the problem that the control scheme
is facing here is the accumulation of mass over the nodes. If all
the energy (represented by the mass) that is injected in the
system was released in avalanches, there will be no large
avalanches. Therefore, one may argue that the ability of the
control scheme can be measured by the ability it has to reduce the
concentration of mass in the system. This kind of information can
be gathered evaluating $\Delta S= S_{\mathrm{2}}-S_{\mathrm{1}}$
where $S_{\mathrm{2}}$ and $S_{\mathrm{1}}$ are respectively the
the Shannon entropy $S(Z)=-\sum_{i=1}^{n}p_{i}\log p_{i}$
evaluated after and before the control intervention and
$p_{i}=z_{i}/\sum_{k=1}^{n} z_{k}$.

It is possible to check whether the entropy change $\Delta S$ is
an useful measure to estimate the ability of the control scheme in
reducing the concentration of mass in the system. For this
purpose, it is necessary to estimate  the probability distribution
$q(\Delta S)$ of success in mass reduction. However, this effect
can be better discussed in terms of the related complementary
probabilities $Q(\Delta S < \Delta S') = \int_0^{\Delta
S'}q(\Delta S) d(\Delta S)$ presented in Fig.2 and its inset.

For the sake of clearness, we divided the values of $\Delta S$ in
two samples corresponding to positive and negative values. While
the main figure shows the complementary distributions for positive
values of $\Delta S$, the inset shows the absolute values of the
negative ones. Comparing the complementary probabilities of the
positive values of $\Delta S$ and also considering the results
presented in Fig.\ref{figura2}, we found that the most efficient
control schemes are more able to reduce the concentration of the
mass of the system. On the other hand, if we compare the negative
values of $\Delta S$ presented in the inset, we note that control
schemes with higher $p_{H}$ increases more the concentration of
mass of the system. This happens since control schemes with higher
$p_{H}$ causes more explosions. However, note that the order of
magnitude of the main figure and the inset are different. Besides,
the random control is the one that causes more increasing of
concentration in the system, since it causes a lot of bad selected
explosions.

Therefore, based on Fig.\ref{figura1} and Fig.\ref{figura2} and
the discursion presented above, it is worth reinforcing that the
ability of the control scheme is related to the ability of
reducing the concentration of the energy of the system. We do not
need to make explosions in all nodes. Since the more important
nodes are selected when we use the H-control, they are enough to
reduce the concentration of energy of the system. On the other
hand, even in the case of the R-control, the concentration of the
energy in the system is also reduced. As we have already
mentioned, if a node is randomly selected, there is good chance
that this node is connected to a high degree and the control
scheme also works. In fact, in both schemes, one may note that the
intervention of the control scheme is not constrained to the
nodes, but only one intervention is sometimes enough to reduce the
concentration of mass in several nodes. We had already mentioned
that another control strategy could be to transport the excess of
mass from the system. Note that although this strategy may be also
interested and investigated somewhere, it has an almost local
effect. Our strategy may affect a large neighborhood of the
network.

In Fig.\ref{figura3} we show results for a ER network built in
order to have the same average degree of the GKK in
Fig.\ref{figura3}. The exponent of the avalanche size distribution
in this ER network was determined to be $\hat \tau=1.52$, using
the interval $s\in[10^{0.3},10^{3.3}]$, which is roughly the value
of the one found in~\cite{bonabeau}. It is also interesting to
compare these results with the ones presented for the GKK
networks. Note that these networks present the same average degree
and also almost similar exponents of avalanche size distribution.
However, for the same $p_{H}$, the number of interventions in the
ER networks are much larger than in the GKK networks. This can be
explained by the bounded heterogeneity of the ER networks, what
requires a larger number of interventions to get similar results.
Indeed, while in the case of the GKK networks the largest
avalanches vanish, the same does not happen to the ER networks.
Finally, though one could expect that H-control and R-control
should present similar results in the case of ER networks, this
does not happen to be verified because of the presence of the
bounded heterogeneity. Due to this particular feature, the
H-control choosing the highest degree nodes is much more efficient
than the R-control.

Fig.\ref{figura4} compares the the probability distribution
function (PDF) of avalanche sizes $p(s)$ of the uncontrolled BTW
(solid symbols) with that of the controlled BTW (hollow symbols)
taking place in the electrical power grid of the western United
States~\cite{watstr98}. The straight line in Fig. \ref{figura3} is
the best fit to the data for this network in the interval
$s\in[10^{0.01},10^{2}]$, which was used to determine the the
exponent of the avalanche size distribution equal to $\hat
\tau=1.45$. It is worthy noting that the range where the PDF can
be approximated by a power law is much smaller than in the two
former cases. We recall that even for GKK or ER networks with
$n=10^4$ nodes, which is roughly the double of the number of the
actual network, the range of power law validity was much larger
than that displayed in Fig.\ref{figura4}. In the inset of this
figure we evaluate the efficiency of the control system by the
ratio $f$ between the number of avalanches of the controlled to
the uncontrolled system showing the good performance of the
control scheme. Although we could expect that the behavior of the
control scheme in this network should perform closer to the ER
network than to the GKK network, it is difficult to prove the
validity of this statement, since the range of validity of the
power law of the avalanche size is much smaller.

\section{Final remarks}We have proposed a control scheme to
reduce the probability of large avalanches of SOC systems on
complex networks. We show that the control scheme works and its
efficiency is based on its ability of reducing the concentration
of mass in the network. This work is an attempt in the direction
of building controlled SOC systems that do not depend on replica
models such as in~\cite{cajand10}.

Unfortunately, we cannot prove optimality of any of these
strategies, since the mathematical model associated to this system
is very complicated being a large set of non-linear coupled
difference equations. In fact, in order to reach optimality, one
should deal with partial removal of the demand from the nodes that
are likely to become critical and consider all the possible order
of triggering the avalanches. Partial removal may work worse than
both strategies. Although partial removal may avoid avalanches
created by the control scheme, it can allow the system to
accumulate energy that in the future can cause larger avalanches.
Therefore, we only intend to show that it is possible to reduce
the size of avalanches on complex networks.

A really interesting agenda of research should be to develop a
control system based on some kind of optimization principle that
could be in some sense reduce the concentration of mass in the
network. Another motivating path would be to apply these ideas in
some real SOC systems.

\section{Acknowledgment} D.O. Cajueiro and R.F.S. Andrade are partially
supported by the Brazilian agency CNPQ. The authors also thank
Professor D. J. Watts for making available the network of the
electrical power grid of the western United States available.

\end{document}